\documentstyle[preprint,aps,epsf]{revtex}
\newcommand{\corr}{\stackrel{\wedge}{=}}
\def\ins#1{}
\def\ind#1{#1}
\def\iem#1{{\em #1\/}}
\def\comment#1{}
\def\eff{{\rm eff}}
\def\ener{\varepsilon}
\def\aut#1{#1}
\def\Tr{{\mbox{Tr\,}}}

\def\Det{{\mbox{Det\,}}}
\def\sfrac#1#2{\raisebox{0.09ex}{\scriptsize${\frac{#1}{#2}}$}}

\newcommand{\sdag}{{\scriptsize \dagger}}
\newcommand{\deltabar}{\,\,{\bar{}\hspace{1pt}\! \!\delta }}

\begin{document}
\draft
\sloppy

\renewcommand{\baselinestretch}{1.3}       %

\title{\vspace{-7.5cm} ~ \hfill {\normalsize FUB-HEP/95}\\[0.5cm]
Classical and Fluctuating Paths \\
in Spaces with Curvature and Torsion
\vspace{-0mm}}
\author{
H.~Kleinert\thanks{
 Email: kleinert@einstein.physik.fu-berlin.de; URL:
http://www.physik.fu-berlin.de/\~{}kleinert; phone/fax: 0049/30/8383034
}}
\address{Institut f\"ur Theoretische Physik,\\
          Freie Universit\"at Berlin,
          Arnimallee 14, D - 14195 Berlin\vspace{-0cm}}
\vspace{-0mm}

\maketitle
\begin{abstract}
\vspace{-6mm}~\\
{\footnotesize This lecture elaborates on a recently discovered
mapping
procedure by which classical orbits and
path integrals for the motion of a point particle
in flat space can be
transformed correctly
into
those in curved space.
This procedure evolved from
well established methods
in the theory of plastic deformations
where crystals with defects are described
mathematically by
applying nonholonomic coordinate
transformations
to ideal crystals.
In the context of
time-sliced path integrals,
there seems to exists a {\em quantum equivalence principle\/}
which determines the measures of fluctating orbits
in spaces with curvature and torsion.
The nonholonomic
transformations
produce a nontrivial
Jacobian in the path measure
which
in a curved space produces
 an additional term
proportional to the curvature scalar
canceling a similar term found earlier
by DeWitt from a naive formulation of Feynman's
time-sliced path
integral.
This cancelation is
important in correctly
describing semiclassically and quantum mechanically
various
systems such as the hydrogen atom, a particle on the surface of a sphere,
and a spinning top. It is also
indispensible for the process of bosonization,
by which Fermi particles
are redescribed in terms of Bose fields.
}
\end{abstract}
\newpage
\section{Introduction}

In 1956, Bryce DeWitt
proposed a path integral formula
for a point particle in
a curved space
using a specific generalization
of Feynman's time-sliced formula
in cartesian coordinates \cite{dw}.
Surprisingly, his amplitude turned out to satisfy a Schr\"odinger equation
different from what had previously been considered as correct
\cite{pod}. In addition to the Laplace-Beltrami
operator for the kinetic term,
his Hamilton operator
contained
an extra effective potential
proportional to the curvature scalar $R$.
At the time of his writing, DeWitt could not think of any
argument to outrule the presence of such an extra term.

Since DeWitt's pioneering work, the time-sliced
path integral in curved spaces
has been reformulated by many people
in a variety of ways \cite{variety}.
The basic problem is the freedom in time slicing
the functional integral. Literature offers prepoint, midpoint,
and postpoint prescriptions which in the Schr\"odinger equation
correspond to different orderings of the momentum operators
$\hat p_\mu$ with respect to the position
variables $q^ \lambda$ in the Hamiltonian operator
$\hat H=g^{\mu \nu}(q)\hat p_\mu \hat p_\nu/2$.
Similar ambiguities are well known
in the theory of stochastic differential equations
where different algorithms have been developed
by Ito and Stratonovich
based on different time discretization procedures \cite{stochasticde}.
In these approaches one may derive
covariant versions
of the Fokker-Planck equation of Brownian motion
in curved spaces.
The mathematical approach to path integrals uses
similar techniques \cite{mathem}.
The inherent
ambiguities can be removed by demanding a
certain form for
Schr\"odinger equation
of the system,
which in curved space
is have the Laplace-Beltrami operator
as an operator for the kinetic energy \cite{pod},
without an
additional curvature scalar.

It has often been repeated that
a Hamiltonian
whose kinetic term
depends on the position variable
has in principle
many different operator versions.
For an
arbitrary model Hamiltonian, this is of course, true.
A specific
physical system, however,
must have a unique Hamilton operator.
If a system has a high symmetry,
it is often possible to
find its correct form
on the basis of group theory.
Recall that in standard textbooks on quantum mechanics
\cite{ll}, a spinning top is quantized by expressing
its Hamiltonian  in terms of the generators of the rotation group,
and quantizing these via the
well-known commutation rules, rather than
canonical variables.
This procedure avoids the ordering problem
by avoiding canonical variables.
 The resulting
Hamilton operator
contains only
a Laplace-Beltrami operator and {\em no extra term\/}
 proportional to $R$.
A spinning top and a particle on the surface of a sphere
are quantized similarly.
This procedure forms the basis of the so called
{\em group quantization\/}
or {\em geometric quantization\/} \cite{geom}
which corresponds to Schr\"odinger equations
containing only the Laplace-Beltrami operator
and no extra curvature terms.

Until recently, geometric quantization was the only
procedure which {\em predicted\/}
the form of the Schr\"odinger equation uniquely
on the basis of symmetry,
with generally accepted results.
Moreover, canonical quantization in flat space is a special case
since it corresponds to a geometric quantization
of the generators of the euclidean group.
Unfortunately, it is quite difficult
to generalize this procedure to
systems in more general geometries without symmetry.
In particular, no earlier approach
makes predictions as to the form of the Schr\"odinger equation
in spaces with curvature and torsion.
In DeWitt's time-sliced approach and its various successors,
Schr\"odiner
equations are found
which contain many possible selections of
scalar combinations of curvature and torsion tensor.
There is a definite need for a
principle capable of predicting the correct
Schr\"odinger equation in such
spaces.

In the context gravity,
this may seem a somewhat
academic question
since nobody has ever
experimentally
observed
a scalar term in the Schr\"odinger equation
of gravitating matter,
for a point particle  in a curved space,
even if torsion is neglected,
and it is not even clear, whether
gravity will generate torsion outside spinning matter.
In other contexts,
however, the problem is not so far from physical reality.
There exist a number of systems which
can be described in  different ways, often gaining
enormously in simplicity by such redescriptions.
A common example is  the
Hamilton operator of a free particle
in euclidean space parametrized in terms of spherical coordinates.
 The operator turns into
 the Laplace-Beltrami operator.

Unfortunately, the use such reparametrizations
is limited.
Since they do not change the geometry of the system,
they are incapble of teaching us
what physics to expect in a space with more
complicated geometries.

There exists, however, an important field in physics where
relations between different geomentries
have been established  successfully for a long time.
This is the physics of plasticity and defects \cite{plastic}.
Defects are described mathematically
by means of what are called {\em nonholonomic\/} coordinate transformations.
In Fig. 1
we show two typical
elementary defects in two dimensions
which can be generated by such
transformations.
It has been understood
a long time ago that, geometrically,
crystals with defects
correspond to spaces with curvature
and torsion \cite{kro}.

In the context of path integrals, such transformation
are of crucial importance. They provided us with
a key
to finding the resolvent of the
most elementary atomic system,
the
hydrogen atom \cite{dk}.
Two such transformations brought it to a harmonic form.
Only recently was recognized
one of these transformation leads to a space
with torsion \cite{PI}.
If DeWitt's
construction rules for a path integral in curved space
are adapted the  associated path integral,
it would produce the wrong atomic spectrum.

The resolution of this puzzle has led to the discovery of
a simple rule for correctly transforming Feynman's time-sliced path integral
formula from its well-known cartesian form
to spaces with curvature and torsion
\cite{ct1,ct2,PI}.
The rule plays the same
fundamental
role in quantum physics
as Einstein's
{\em equivalence principle\/}
does within classical physics,
where it governs the form of
the equations of motion in curved spaces.
It
is therefore called {\em  quantum equivalence principle\/} (QEP) \cite{PI}.

The crucial place where this principle makes a nontrivial statement
is in the measure of the
path integral. The nonholonomic nature of the
differential coordinate transformation gives rise to
an additional term with respect to the naive
DeWitt measure, and this
 cancels
 precisely the
bothersome term proportional to $R$ found.
In the presence of torsion, the quantum equivalence principle
{\em predicts\/} a simple Schr\"odinger
equation.
No other approach
has done this before,

It is the purpose of these lectures
to demonstarte the power of the
new
quantum equivalence principle
and to discuss wits consequences
also at the
classical level,
where
the familiar action principle breaks down and requires
an important modification \cite{fk1,PI}. The geometric reason for this is
that infinitesimal variations can no longer
be taken as closed curves; they
possess a defect analogous to the Burgers vector
in crystal physics.
This surprizing
result
has been verified by deriving the Euler equations
for the motion of a spinning top
from an action principle formulated
{\em within\/}
the body-fixed reference frame, where the geometry of the
nonholonomic coordinates possesses torsion \cite{fk2}.

\section
[Classical Motion of a Mass Point in a Space with Torsion~]
{Classical Motion of a Mass Point in a Space with Torsion}
\index{gravitational field, classical motion in}
\index{classical motion in gravitational field}
We begin by recalling that
Einstein formulated the
rules for finding the classical laws of motion in a gravitational
field
on the basis of his famous
 \ind{equivalence principle}.
He assumed the space to be free of torsion since
otherwise his
geometric priciple
was not able to determine the
classical equations of motion uniquely.
Since our nonholonomic mapping principle is free of
this problem,
we
do not need to rescrict the geometry in this way.
The correctness of the resulting laws of motion
is exemplified by several physical systems
with well-known experimental properties. Basis for
these ``experimental verifications" will be the fact
that classical equations of motion are invariant
under nonholonomic coordinate transformations.
Since it is well known that such transformations introduce curvature
 and torsion into
a parameter space,
such redescriptions of standard mechanical systems
provide us with sample systems
in
general metric-affine spaces.

To be as specific and  as simple as possible,
we first formulate the theory for a
nonrelativistic massive point particle in a general metric-affine space.
The entire discussion may easily be
extended to relativistic
particles in spacetime.

\subsection{Equations of Motion}

Consider the action of the
particle along the orbit ${\bf x}(t)$ in a flat
space
parametrized with rectilinear, Cartesian coordinates:
\begin{equation} \label{10.1}
{\cal A} = \int_{t_a}^{t_b} dt \frac{M}{2}( {\dot{x}^i})^2,
\;\;\;\; i=1,2,3.
\end{equation}
It may be transformed to curvilinear coordinates $q^\mu, ~\mu =1,2,3$, via some
functions
\begin{equation} \label{10.2}
x^i = x^i(q),
\end{equation}
leading to
\begin{equation} \label{10.3}
{\cal A} = \int_{t_a}^{t_b} dt \frac{M}{2} g_{\mu\nu}(q) \dot{q}^\mu
\dot{q}^\nu,
\end{equation}
where
\begin{equation} \label{10.4}
g_{\mu\nu}(q) = \partial_\mu x^i (q) \partial_\nu x^i (q)
\end{equation}
is the {\iem{induced metric}} for the curvilinear coordinates.
Repeated indices are understood to be summed over, as usual.

The length
of the orbit in the flat space is given by
\begin{equation} \label{10.5}
l = \int_{t_a}^{t_b} dt \sqrt{g_{\mu\nu} (q) \dot{q}^\mu \dot{q}^\nu}.
\end{equation}
Both the action (\ref{10.3}) and the length (\ref{10.5})
are invariant under arbitrary \iem{reparametrizations of space} $q^ \mu
\rightarrow q'{}^ \mu $.

Einstein's equivalence
principle\index{Einstein's equivalence principle}\index{equivalence principle}
amounts to the postulate that
the transformed action (\ref{10.3}) describes directly the motion
of the particle in the presence of a gravitational field
caused by other masses.
The  forces caused by
 the  field are all a result of
 the geometric properties
of the metric tensor.

The equations of motion are obtained by
extremizing the action in Eq.~(\ref{10.3})
with the result
\begin{equation} \label{10.6}
\partial_t(g_{\mu\nu} \dot{q}^\nu)-\frac{1}{2}\partial _\mu
g_{\lambda  \nu }\dot q^\lambda \dot q^\nu  = g_{\mu\nu} \ddot{q}^\nu
+ \bar{\Gamma}_{\lambda\nu\mu} \dot{q}^\lambda \dot{q}^\nu = 0.
\end{equation}
Here\vspace{-0mm}
\begin{equation} \label{10.7}
\bar{\Gamma}_{\lambda\nu\mu} \equiv \frac{1}{2} (\partial_\lambda
g_{\nu\mu} + \partial_\nu g_{\lambda\mu} -
\partial_\mu g_{\lambda\nu})
\end{equation}
is the {\iem{Riemann connection}} or {\iem{Christoffel symbol}}
of the {\em first kind\/}. Defining also the Christoffel
symbol of the {\em second kind\/}
\begin{equation} \label{10.8}
\bar{\Gamma}_{\lambda\nu}^{\;\;\;\;\mu} \equiv g^{\mu\sigma}
\bar{\Gamma}_{\lambda\nu\sigma}  ,
\end{equation}
we can write\vspace{-0mm}
\begin{equation} \label{10.9}
\ddot{q}^\mu + {{{{\bar{\Gamma}}_{\lambda\nu}}}}^{\;\;\;\;\mu}
\dot{q}^\lambda \dot{q}^\nu = 0.
\end{equation}
The solutions of these equations are the classical orbits. They coincide
with the extrema of the length of a curve $l$ in (\ref{10.5}).
Thus, in a curved space, classical orbits are the shortest curves,
called {\iem{geodesics}}.

The same equations can also be obtained directly by
transforming the equation of motion from\vspace{-0mm}
\begin{equation} \label{10.10}
\ddot{x}^i = 0
\end{equation}
to curvilinear coordinates $q^\mu$, which gives
\begin{equation} \label{10.11}
\ddot{x}^i = \frac{\partial x^i}{\partial q^\mu} \ddot{q}^\mu
+ \frac{\partial^2 x^i}{\partial q^\lambda \partial q^\nu}
\dot{q}^\lambda\dot{q}^\nu = 0.
\end{equation}
At this place it is useful to employ the
 so-called {\iem{basis triads}}\vspace{-0mm}
\begin{equation} \label{10.12}
{e^i}_\mu (q) \equiv \frac{\partial x^i}{\partial q^\mu}
\end{equation}
and the {\iem{reciprocal basis triads}}
\begin{equation} \label{10.13}
{e_i}^\mu (q) \equiv \frac{\partial q^\mu}{\partial x^i},
\end{equation}
which satisfy the orthogonality and completeness relations
\begin{eqnarray} \label{10.14}
{e_i}^\mu {e^i}_\nu & = & {\delta^\mu}_\nu,\\
 {e_i}^\mu {e^j}_\mu & = & {\delta_i}^{j}.
\end{eqnarray}
The \ind{induced metric} can then be written as
\begin{equation} \label{10.16}
g_{\mu\nu} (q) =
{e^i}_\mu (q) {e^i}_\nu (q).
\end{equation}
Labeling Cartesian coordinates,
upper and
lower indices $i$ are the same. The indices
 $\mu, \nu$ of the curvilinear coordinates, on the other hand,
can be lowered only by contraction with the metric $g_{\mu\nu}$
or raised with the inverse metric $g^{\mu\nu}\equiv(g_{\mu\nu})^{-1}$.
Using the basis triads,
Eq.~(\ref{10.11}) can be rewritten as
\begin{equation} \label{10.17}
\frac{d}{dt} ({e^i}_\mu \dot{q}^\mu) = {e^i}_\mu
\ddot{q}^\mu
+ {e^i}_{\mu,\nu} \dot{q}^\mu \dot{q}^\nu = 0,\nonumber
\end{equation}
or as
\begin{equation} \label{10.18}
\ddot{q}^\mu + {e_i}^\mu {e^i}_{\kappa,\lambda}
\dot{q}^\kappa \dot{q}^\lambda=0.
\end{equation}
The subscript $ \lambda $ separated by a comma denotes the
partial derivative
$\partial_\lambda = \partial/\partial q^\lambda$ , i.e.,
$f_{,\lambda} \equiv \partial_\lambda f$.
The quantity in front of $\dot{q}^\kappa \dot{q}^\lambda$ is
called the {\iem{affine connection}}:
\begin{equation} \label{10.19}
{\Gamma_{\lambda\kappa}}^{\mu} = {e_i}^\mu  {e^i}_{\kappa ,\lambda}.
\end{equation}
Due to (\ref{10.14}), it can also be written as
\begin{equation} \label{10.20}
{\Gamma_{\lambda\kappa}}^\mu = - {e^i}_\kappa {{{e_i}^\mu}}_{,\lambda}.
\end{equation}
Thus we arrive at the transformed flat-space equation of motion
\begin{equation} \label{10.21}
\ddot{q}^\mu + {\Gamma_{\kappa\lambda}}^\mu \dot{q}^\kappa
\dot{q}^\lambda = 0.
\end{equation}
The solutions of this  equation
are called the {\iem{straightest lines}} or {\iem{autoparallels}}.

If the coordinate transformation $x^i(q)$ is
smooth and single-valued, it is integrable, i.e.,
the derivatives commute as required
by Schwarz's \ind{integrability condition}
\begin{equation} \label{10.22}
(\partial_\lambda \partial_\kappa - \partial_\kappa \partial_\lambda )
x^i(q) = 0.
\end{equation}
Then the triads satisfy the identity
\begin{equation} \label{10.23}
{e^i_{\kappa,\lambda}} = {e^i_{\lambda , \kappa}},
\end{equation}
implying that the connection  ${\Gamma_{\mu\nu}}^\lambda$ is symmetric in the
lower indices.
In this case it
coincides with the \ind{Riemann connection}, the \ind{Christoffel symbol}
$\bar{\Gamma}_{\mu\nu}^{\;\;\;\;\lambda}$. This follows
immediately after inserting
$g_{\mu\nu}(q)={e^i}_\mu(q) {e^i}_\nu (q)$
into (\ref{10.7}) and working out all derivatives using
(\ref{10.23}).
Thus, for a space with  curvilinear coordinates $q^\mu$ which can be
reached by an
integrable coordinate transformation from a flat space, the autoparallels
coincide with the geodesics.

\subsection[Nonholonomic Mapping to Spaces with Torsion]
{Nonholonomic Mapping to Spaces with Torsion}
It is possible to map the $x$-space locally into a $q$-space
via an infinitesimal transformation
\begin{equation}
dx^i=e^i{}_ \mu (q)dq^ \mu ,
\label{10.diffi}\end{equation}
with coefficient functions
$e^i{}_ \mu (q)$ which are
not integrable in the sense of  Eq.~(\ref{10.22}), i.e.,
\begin{equation}
\partial _ \mu e^i{}_ \nu (q)
-\partial _ \nu e^i{}_ \mu (q)\neq0.
\label{5.swr}\end{equation}
Such a mapping  will be called \iem{nonholonomic}.
It does not lead
to a
single-valued function $x^i(q)$.
Nevertheless, we shall write (\ref{5.swr})  in analogy to
(\ref{10.22})
as
\begin{equation} \label{10.22x}
(\partial_\lambda \partial_\kappa - \partial_\kappa \partial_\lambda )
x^i(q) \neq 0,
\end{equation}
since this equation involves only the differential $dx^i$.
Our departure from mathematical conventions will not cause any
problems.

{}From Eq.~(\ref{5.swr}) we see that
the image space of a
nonholonomic mapping
carries torsion. The
connection
$ {\Gamma_{\lambda\kappa}}^{\mu} = {e_i}^\mu  {e^i}_{\kappa ,\lambda}$
has a nonzero antisymmetric part, called the \iem{torsion tensor}:\footnote{Our
notation for
the geometric quantities in
spaces with curvature and torsion is the same as in
\aut{J.A.~Schouten}, {\em Ricci Calculus\/}, Springer, Berlin, 1954.}
\begin{equation} \label{10.24}
{S_{\lambda\kappa}}^\mu  = \frac{1}{2} ({\Gamma_{\lambda\kappa}}^\mu
 -{\Gamma_{\kappa\lambda}}^\mu).
\end{equation}
In contrast to ${\Gamma_{\lambda\kappa}}^\mu$,
the antisymmetric part ${S_{\lambda\kappa}}^\mu$
is a proper tensor under general coordinate transformations.
The  contracted tensor
\begin{equation} \label{10.25}
S_\mu \equiv {S_{\mu\lambda}}^\lambda
\end{equation}
transforms like a vector, whereas the contracted connection
$\Gamma_\mu \equiv {\Gamma_{\mu\nu}}^\nu $
does not.
Even though ${\Gamma_{\mu\nu}}^\lambda $ is not a tensor,
we shall freely lower and raise its indices using contractions with
the metric or the inverse metric, respectively: ${\Gamma^\mu}_{\nu}{}^\lambda
\equiv
g^{\mu\kappa} {\Gamma_{\kappa\nu}}^\lambda $, ${\Gamma_\mu}^{\nu}{}^\lambda
\equiv
g^{\nu \kappa} {\Gamma_{\mu \kappa}}^\lambda $, $\Gamma_{\mu\nu\lambda} \equiv
g_{\lambda \kappa} {\Gamma_{\mu \nu}}^\kappa $. The same thing will
be done with
$ \bar{\Gamma}_{\mu \nu }{}^\lambda $.

In the presence of torsion, the connection
is no longer equal to the Christoffel symbol.
In fact,
by rewriting $\Gamma _{\mu \nu \lambda }=e_{i\lambda }\partial _\mu e^i{}_\nu $
trivially as
\begin{eqnarray} \label{10.26}
&&\!\!\!\!\!\!\!\!\!\!\!\!\!\Gamma _{\mu \nu \lambda }=\frac{1}{2}\left\{
e_{i\lambda }\partial  _\mu e^i{}_\nu
+\partial _\mu e_{i\lambda } e^i{}_\nu
+e_{i\mu      }\partial  _\nu  e^i{}_\lambda
+\partial _\nu e_{i\mu } e^i{}_\lambda
-e_{i\mu     }\partial  _\lambda  e^i{}_\nu
-\partial _\lambda  e_{i\mu  } e^i{}_\nu\right\} \nonumber \\
&&\!\!\!\!\!\!\!\!+\frac{1}{2}\left\{
\left[  e_{i\lambda }\partial  _\mu e^i{}_\nu -e_{i\lambda }\partial  _\nu
e^i{}_\mu\right]
-\left[ e_{i\mu }\partial  _\nu e^i{}_\lambda  -e_{i\mu }\partial  _\lambda
e^i{}_\nu\right]
+\left[ e_{i\nu  }\partial  _\lambda  e^i{}_\mu -e_{i\nu  }\partial  _\mu
e^i{}_\lambda  \right] \right\} \nonumber
\end{eqnarray}
and using
${e^i}_\mu(q) {e^i}_\nu (q)=g_{\mu\nu}(q)$,
we find the decomposition
\begin{equation} \label{10.26b}  
{\Gamma_{\mu\nu}}^\lambda ={ \bar\Gamma }_{\mu\nu }^{\;\;\;\;\lambda} +
{K_{\mu\nu}}^\lambda,
\end{equation}
where the combination of torsion tensors
\begin{equation} \label{10.27a}
K_{\mu\nu\lambda} \equiv S_{\mu\nu\lambda} -
S_{\nu\lambda\mu} + S_{\lambda\mu\nu}
\end{equation}
is called the {\iem{contortion tensor}}.
It is antisymmetric in the last two indices so that
\begin{equation} \label{10.27}
\Gamma _{\mu \nu }{}^{\nu }=\bar \Gamma _{\mu \nu }{}^{\nu }.
\end{equation}

In the presence of torsion, the shortest and straightest lines are no
longer equal. Since the two types of lines play geometrically an
equally favored role, the question arises as to which of them describes
the correct classical particle orbits. The answer will be given at the
end of this section.

The main effect of matter in Einstein's theory
of gravitation
manifests itself in the violation of the
 \ind{integrability} condition for the derivative of
the coordinate transformation $x^i(q)$, namely,
\begin{equation} \label{10.28}
(\partial_\mu \partial_\nu - \partial_\nu \partial_\mu)
\partial_\lambda x^i (q) \neq 0.
\end{equation}
A transformation for which $x^i (q)$ itself is integrable, while the
first derivatives $\partial_\mu x^i (q)= {e^i}_\mu $ are not,
carries a flat-space region into a purely curved one.
The quantity which
records the nonintegrability is the {\iem{Cartan curvature tensor}}
\begin{equation} \label{10.29}
{R_{\mu\nu\lambda}}^\kappa = {e_i}^\kappa
(\partial_\mu \partial_\nu - \partial_\nu \partial_\mu) {e^i}_\lambda.
\end{equation}
Working out the derivatives using (\ref{10.19}) we see that
${R_{\mu\nu\lambda}}^\kappa$ can be written as a covariant curl
of the connection,
\begin{equation} \label{10.30}
{R_{\mu\nu\lambda}}^\kappa = \partial_\mu {\Gamma_{\nu\lambda}}^\kappa
- \partial_{\nu}{\Gamma_{\mu\lambda}}^\kappa
- [\Gamma_\mu , \Gamma_\nu{]_\lambda}^\kappa.
\end{equation}
In the last term we have used a matrix notation for the connection.
The tensor components
  ${\Gamma_{\mu\lambda}}^\kappa$ are viewed as matrix elements
$(\Gamma_\mu{)_\lambda}^\kappa$, so that we can use the matrix
commutator
\begin{equation} \label{10.31}
[\Gamma_\mu , \Gamma_\nu{]_\lambda}^\kappa \equiv
(\Gamma_\mu \Gamma_\nu - \Gamma_\nu \Gamma_\mu{)_\lambda}^\kappa
= {\Gamma_{\mu\lambda}}^\sigma {\Gamma_{\nu\sigma}}^\kappa
- {\Gamma_{\nu\lambda}}^\sigma {\Gamma_{\mu\sigma}}^\kappa.
\end{equation}

Einstein's original theory of gravity
assumes the absence of
 torsion. The space properties are
 completely specified by the
{\iem{Riemann curvature tensor}}
formed from the \ind{Riemann connection} (the \ind{Christoffel symbol})
\begin{equation} \label{10.32}
{{\bar{R}}_{\mu\nu\lambda}}^{\;\;\;\;\;\;\kappa} = \partial_\mu
{ \bar{\Gamma }}_{\nu\lambda}^ {\;\;\;\;\kappa} - \partial_\nu
{\bar{\Gamma}}_{\mu\lambda}^{\;\;\;\;\kappa}
- [\bar{\Gamma}_\mu , \bar{\Gamma}_\nu {]_\lambda}^\kappa.
\end{equation}
The relation between the two curvature tensors is
\begin{equation} \label{10.33}
{R_{\mu\nu\lambda}}^\kappa = { {\bar{R}}_{\mu\nu\lambda}}^{\;\;\;\;\;\;\kappa}
+ \bar{D}_\mu {K_{\nu\lambda}}^\kappa
- \bar{D}_\nu {K_{\mu\lambda}}^\kappa - {[K_\mu , K_\nu ]_\lambda}^\kappa
{}.
\end{equation}
In the last term, the ${K_{\mu\lambda}}^\kappa$'s are viewed as matrices
$(K_\mu
{)_\lambda}^\kappa$. The symbols $\bar{D}_\mu$ denote the
{\iem{covariant derivatives}}
formed with the Christoffel symbol.
Covariant derivatives act like ordinary derivatives
if they are applied to a scalar field.
 When applied to a vector field,
 they act as follows:
\begin{eqnarray} \bar{D}_\mu  v_\nu &  \equiv  & \partial_\mu v_\nu -
{{\bar{\Gamma}}_{\mu\nu}}^{\;\;\;\;\lambda}
v_\lambda, \nonumber\\
\bar{D}_\mu v^\nu &  \equiv  & \partial_\mu v^\nu +
{\bar\Gamma_{\mu\lambda}}^{\;\;\;\;\nu}
v^\lambda.   \label{10.34}
\end{eqnarray}
The effect upon a tensor
field is the generalization of this;
 every index receives a corresponding
additive $\bar{\Gamma}$ contribution.

In the presence of torsion,  there exists another covariant derivative
formed with the affine connection $ {\Gamma_{\mu\nu}}^\lambda$
rather than the Christoffel symbol   which acts
upon a vector field as
\begin{eqnarray}D_\mu v_\nu &  \equiv & \partial_\mu v_\nu -
{\Gamma_{\mu\nu}}^\lambda
v_\lambda , \nonumber\\
D_\mu  v^\nu &  \equiv  & \partial_\mu v^\nu + {\Gamma_{\mu\lambda}}^\nu
v^\lambda.     \label{10.35}
\end{eqnarray}
This will be of use later.

{}From either of the two curvature tensors, ${R_{\mu\nu\lambda}}^\kappa$ and
${{\bar{R}}_{\mu\nu\lambda}}^{\;\;\;\;\;\;\kappa}$, one
can form the once-contracted tensors of rank 2, the {\iem{Ricci tensor}}
\begin{equation} \label{10.36}
R_{\nu\lambda} = {R_{\mu\nu\lambda}}^\mu,
\end{equation}
and the {\iem{curvature scalar}}
\begin{equation} \label{10.37}
R = g^{\nu\lambda} R_{\nu\lambda}.
\end{equation}
The  celebrated \ind{Einstein equation}
 for the gravitational
field postulates that the tensor
\begin{equation} \label{10.38}
G_{\mu\nu} \equiv  R_{\mu\nu} - \frac{1}{2} g_{\mu\nu} R,
\end{equation}
the so-called {\iem{Einstein tensor}}, is proportional to
the \ind{symmetric energy-momentum tensor} of all matter fields.
This postulate was made only for spaces with no torsion, in which case
$R_{\mu\nu} = \bar{R}_{\mu\nu}$ and $ R_{\mu \nu },\,G_{\mu \nu }$  are
both symmetric. As mentioned before, it is not yet
clear how Einstein's
field equations should be generalized in the presence of torsion
since the experimental consequences are
as yet too small to be
observed.
In this paper, we are not concerned with
the generation of curvature and torsion but only with
their consequences upon the motion of point particles.

The generation of defects illustrated in Fig. 1
provides us with
two simple examples
for nonholonomic mappings which show us
in which way these mappings are capable of generating a space with
curvature and torsion
from a euclidean space.\index{nonintegrable mappings}\index{nonholonomic
mappings}
The reader not familiar with this subject
is advised to  consult the standard literature
on this subject quoted in Ref. \cite{plastic,kro}.

Consider first the upper example
in which a dislocation is generated,
characterized by
 a
missing or additional layer of atoms (see Fig.~10.1).
In two dimensions, it may be described differentially
by the transformation
\begin{equation} \label{10.39}
dx^i =
\left\{
\begin{array}{ll}
dq^1 & ~~~\mbox{for $i = 1$,} \\
dq^2  +  \ener \partial_\mu \phi(q) dq^\mu & ~~~\mbox{for $i = 2$,}
\end{array}\right.
\end{equation}
with the multi-valued function
\begin{equation} \label{10.40}
\phi(q)\equiv \arctan (q^2/q^1).
\end{equation}
The triads reduce to dyads, with the components
\begin{eqnarray} \label{10.41}
{e^1}_\mu & = & {\delta^1}_\mu~~ ,   \nonumber\\
{e^2}_\mu & = & {\delta^2}_\mu + \epsilon \partial_\mu \phi(q)~~,
\end{eqnarray}
and the  \ind{torsion tensor} has
the components
\begin{equation} \label{10.42a}
{e^1}_\lambda {S_{\mu\nu}}^\lambda = 0 ,\;\;\;\;\;
{e^2}_\lambda {S_{\mu\nu}}^\lambda = \frac{\epsilon}{2} (\partial_\mu
\partial_\nu - \partial_\nu \partial_\mu) \phi.
\end{equation}
If we differentiate (\ref{10.40}) formally,
we find
$(\partial_\mu \partial_\nu - \partial_\nu \partial_\mu)
\phi \equiv 0$. This, however, is incorrect at the origin.
Using Stokes'~ theorem we see that
\begin{equation} \label{10.43}
\int d^2 q (\partial_1 \partial_2- \partial_2 \partial_1)
\phi = \oint dq^\mu \partial_\mu \phi = \oint d \phi = 2\pi
\end{equation}
for any closed circuit around the origin,
implying that there is a $\delta$-function singularity at the
origin with
\begin{equation} \label{10.44}
{e^2}_\lambda {S_{12}}^\lambda = \frac{\epsilon}{2} 2\pi
\delta^{(2)} (q).
\end{equation}
By a linear superposition of
such mappings we can generate an arbitrary torsion in the $q$-space.
The mapping introduces no curvature.
When encircling a dislocation
along
a closed path
 $C$,
its counter image
$C'$
in the ideal crystal
does not form a
closed path.
The \ind{closure failure}
is called the \iem{Burgers vector}
\begin{equation}
b^ i  \equiv
\oint_{C'} dx^i
=\oint_{C} dq^ \mu e^i{}_ \mu .
\label{10.burg1}\end{equation}
It specifies the direction and thickness
of the layer of additional atoms.
With the help of Stokes' theorem,
it is seen to measure the torsion
contained in any surface  $S$ spanned by $C$:
\begin{eqnarray}
b^i& =&
\oint_Sd^2s^{ \mu   \nu  } \partial  _ \mu  e^i{}_  \nu
=\oint_Sd^2s^{ \mu  \nu }S_{ \mu  \nu }{}^  \lambda ,
\label{10.encl1}\end{eqnarray}
where
$
d^2s^{ \mu  \nu }=-
d^2s^{ \nu  \mu }
$
is the projection of an oriented
infinitesimal area element onto the plane $ \mu  \nu $.
The above example has the Burgers vector
\begin{equation}
b^ i =(0, \epsilon ).
\label{}\end{equation}

A corresponding \ind{closure failure} appears when
 mapping a
closed contour $C$ in the ideal crystal
into a crystal containing
a dislocation. This defines a Burgers vector:
\begin{equation}
b^ \mu  \equiv
\oint_{C'} dq^ \mu
=\oint_{C} dx^ie_i{}^ \mu .
\label{10.burg}\end{equation}
By
Stokes' theorem, this becomes a surface integral
\begin{eqnarray}
b^ \mu& =&
\oint_{S}d^2s^{ij} \partial  _i e_j{}^  \mu
=\oint_Sd^2s^{ij}e_i{}^  \nu
\partial _  \nu   e_j{}^  \mu \nonumber \\
&=&-\oint_Sd^2s^{ij}e_i{}^   \nu   e_j{}^  \lambda   S_{ \nu  \lambda   }{}^
\mu ,
\label{10.encl}\end{eqnarray}
the
last step following from (\ref{10.20}).

The second example is the nonholonomic mapping
in the lower part of Fig. 1 generating
a disclination
which corresponds to
 an entire section of angle $\alpha$ missing in an ideal
atomic array.
For an infinitesimal angel $ \alpha$, this
may be described, in two dimensions, by the differential mapping
\begin{equation} \label{10.45}
x^{i } = \delta ^i{}_ \mu [q^ \mu +
 \Omega
 \epsilon ^{ \mu}{}_  \nu q ^\nu\phi(q)],
\end{equation}
with the multi-valued function (\ref{10.40}).
The symbol $ \epsilon _{ \mu  \nu } $ denotes the
antisymmetric Levi-Civit\`a
tensor.
The transformed metric
\begin{equation} \label{10.46}
g_{ \mu  \nu }=
 \delta _{ \mu  \nu }- \frac{2 \Omega  }{q^ \sigma  q_ \sigma  }
 \epsilon _{ \mu  \nu } \epsilon ^ { \mu}{}_{  \lambda }
\epsilon ^{ \nu}{}_{\kappa }
q ^ \lambda q ^ \kappa .
\end{equation}
is single-valued and has commuting derivatives.
The torsion tensor  vanishes since
$(\partial _ 1\partial _ 2 -\partial _ 2 \partial _ 1)x^{1,2}$
is proportional to $q ^{2,1}  \delta ^{(2)}(q)=0$.
The local
rotation
field
$ \omega (q)\equiv \sfrac{1}{2}( \partial _1x^2-\partial _2x^1)$,
on the other hand,
is equal to the multi-valued function
$- \Omega \phi(q)$,
thus having the
 noncommuting derivatives:
\begin{equation}
(\partial _1\partial _2-\partial _2\partial _1)
\omega (q)=- 2\pi \Omega  \delta ^{(2)}(q).
\label{}\end{equation}
To lowest order in $ \Omega $, this determines
the curvature tensor,
which in two dimensions posses only one independent component, for instance
$R_{1212}$.
Using the fact that $g_{ \mu  \nu }$ has commuting derivatives,
$R_{1212} $ can be written as
\begin{equation}
R_{ 1212 }=(\partial _ 1 \partial _ 2 -\partial _ 2 \partial _ 1 )
 \omega (q) .
\label{}\end{equation}

\subsection{New Equivalence Principle}

In classical mechanics, many dynamical problems
are solved with the help of nonholonomic transformations.
Equations of motion are differential equations
which remain valid if transformed differentially
to new coordinates, even if the transformation is not integrable
in the Schwarz sense.
Thus we \iem{postulate} that the correct
equation of motion of
point particles in
a space with curvature and torsion
are the images of the equation of motion in a flat space.
The
equations
(\ref{10.21}) for the autoparallels yield therefore the correct
trajectories of spinless point particles in a space with curvature
and torsion.

This postulate is based on our knowledge of the motion
of many physical systems. Important examples are the Coulomb system
which is discussed in detail in Chapter~13
of Ref.~\cite{PI}, the
spinning top in the body-fixed reference system \cite{fk2},
and a bosonized version
of a path integral of a
set of Fermi fields \cite{boz}.
Thus the postulate has a good chance of being true, and
will henceforth
be referred to as a \iem{new equivalence principle}.

\subsection[Classical Action Principle for Spaces with Curvature and Torsion]
{Classical Action Principle for Spaces \\ with Curvature and Torsion}

Before setting up a path integral
for the time evolution amplitude
we must find an action principle for
the classical motion
of a
spinless point particle
in a space with curvature and torsion, i.e., the movement along
autoparallel trajectories.
This is a nontrivial task since
autoparallels must emerge
as the extremals of an action  (\ref{10.3})
involving
only the metric
tensor $g_{ \mu  \nu }$. The action is independent of the torsion and
carries only information
on the Riemann part of the space geometry.
Torsion can therefore enter the equations of motion only via
some novel feature of the
variation procedure.
Since we know how to perform variations of an action in the euclidean
$x$-space,
we deduce
the correct procedure in the general \ind{metric-affine space}
by transferring
 the
  variations $\delta x^i(t)$
under the nonholonomic mapping
\begin{equation}
\dot q ^\mu=e_i{}^ \mu(q) \dot x^i
\label{10.diffbez}\end{equation}
into the $q^ \mu $-space. Their
 images
 are quite different
from ordinary variations as illustrated in Fig.~\ref{10.pnonh}(a).
The variations  of the Cartesian coordinates $\delta x^i(t)$
are done
at fixed end points  of
the paths. Thus
they form \iem{closed paths} in  the $x$-space.
Their images, however,
 lie in a space with defects
and thus  possess a \ind{closure failure}
indicating the amount of torsion
introduced by the mapping.
This property will be emphasized by writing the images
 $\deltabar q^\mu (t)$ and calling them {\iem{nonholonomic variations}}.

Let  us calculate them explicitly. The paths in the two spaces
   are related by the integral equation
\begin{equation}
    q^\mu (t) = q^\mu (t_a) + \int^{t}_{t_a}
     dt'  e_i{}^\mu (q(t')) \dot{x}^i(t') .
\label{10.pr2}\end{equation}
%
For two neighboring paths in $x$-space
differing from each other by a
variation $ \delta x^i(t)$,
Eq.~(\ref{10.pr2}) determines the
nonholonomic variation
$\deltabar q^\mu (t)$:
\begin{equation}
 \deltabar    q^\mu (t) =  \int^{t}_{t_a}   dt'
      \deltabar [  e_i{}^\mu (q(t')) \dot{x}^i(t')].
\label{10.pr2p}\end{equation}
A comparison with (\ref{10.diffbez}) shows that
the variations $ \deltabar q^ \mu  $ and the time derivative
of $q^ \mu$
are independent of each other
\begin{equation}
 \deltabar \dot{q}^\mu (t) = \frac{d}{dt} \deltabar q^\mu (t),
\label{10.pdelta}\end{equation}
just as for ordinary variations $ \delta x^i$.

Let us introduce an \iem{auxiliary holonomic variation}s
in
$q$-space:
\begin{equation}
  \delta q^\mu  \equiv e_i{}^\mu (q) \delta x^i.
\label{10.delq}\end{equation}
In contrast to $\deltabar q^\mu (t)$,
these vanish at the endpoints,
\begin{equation}
\delta q(t_a)=
 \delta q(t_b)=0,
\label{10.versch}\end{equation}
i.e., they form closed paths with the unvaried
orbits.

Using (\ref{10.delq})
we derive from (\ref{10.pr2p}) the relation
\begin{eqnarray}
 \frac{d}{dt}\deltabar    q^\mu (t) &=&
      \deltabar  e_i{}^\mu (q(t)) \dot{x}^i(t)
      +  e_i{}^\mu (q(t))\deltabar \dot x^i(t)\nonumber \\
     &=& \deltabar  e_i{}^\mu (q(t)) \dot{x}^i(t)
      +  e_i{}^\mu (q(t)) \frac{d}{dt}[e^i{}_ \nu (t) \delta q^ \nu (t)].
\label{}\end{eqnarray}
After inserting
\begin{equation}
\deltabar  e_i{}^\mu (q)= -\Gamma_{ \lambda  \nu }{}^ \mu \deltabar q^ \lambda
e_i{}^ \nu  ,
{}~~~~~
\frac{d}{dt } e^i{}_\nu (q)= \Gamma_{ \lambda   \nu  }{}^ \mu  \dot q^ \lambda
e^i{}_\mu,
\label{}\end{equation}
this becomes
\begin{equation}
\!\!\!\!\!\!\!\!\frac{d}{dt} \deltabar    q^\mu (t) =  -\Gamma_{ \lambda  \nu
}{}^ \mu \deltabar q^ \lambda \dot q^ \nu
      +
\Gamma_{ \lambda    \nu  }{}^ \mu  \dot q^ \lambda  \delta q^ \nu
+ \frac{d}{dt} \delta  q ^\mu           .
\label{10.prneu}\end{equation}
It is useful to introduce the difference
between the nonholonomic
variation
$ \deltabar q^\mu $
and the auxiliary holonomic variation $ \delta q^ \mu $:
\begin{equation}
   \deltabar b ^\mu\equiv \deltabar q^\mu -\delta q^\mu.
\label{10.deldel}\end{equation}
Then we can rewrite
(\ref{10.prneu})
as a first-order
differential equation for $\deltabar b^ \mu $:
\begin{equation}
   \frac{d}{dt} \deltabar b^ \mu  = -
            \Gamma _{\lambda \nu }{}^\mu  \deltabar b ^\lambda
\dot{q}^\nu
             + 2S _{ \lambda \nu  }{}^\mu
           \dot{q}^ \lambda      \delta q^ \nu  .
\label{10.p83}\end{equation}
After introducing the matrices
\begin{eqnarray}
 \label{10.p15}
G {} ^\mu (t){}_ \lambda   \equiv   \Gamma _{ \lambda  \nu }{}^\mu
               (q(t))\dot{q}^\nu (t)
\end{eqnarray}
and
\begin{eqnarray}
    \Sigma  ^\mu{}_\nu (t) \equiv  2   S_{\lambda \nu } {}^\mu
                 (q(t)) \dot{q}^ \lambda  (t)
                ,
\label{10.p2equ}\end{eqnarray}
equation (\ref{10.p83}) can be written as a vector differential
equation:
\begin{equation}
   \frac{d}{dt} \deltabar b  = - G\deltabar b
            +  \Sigma (t)\,\delta q^ \nu  (t).
\label{10.p83p}\end{equation}
This is solved by
\begin{eqnarray}
   \deltabar b (t) = \int^{t}_{t_a} dt'
          U({t,t'})~ \Sigma  (t')~\delta q  (t'),
\label{10.pdel}\end{eqnarray}
with the matrix
\begin{equation}
  U({t,t'}) = T \exp \left[ -\int^{t}_{t'} dt'' G(t'')\right].
\label{10.pum}\end{equation}
In the absence
of torsion,
 $\Sigma (t)
 $ vanishes identically and $\deltabar b (t) \equiv 0$,
and the
variations
 $\deltabar q^\mu (t)$
coincide with the
holonomic
 $\delta  q^\mu (t)$ [see Fig.~\ref{10.pnonh}(b)].
In a space with torsion,
the
variations
 $\deltabar q^\mu (t)$ and $\delta  q^\mu (t)$
are different
from each other
[see
 Fig.~\ref{10.pnonh}(c)].

Under an arbitrary nonholonomic
variation   $\deltabar q^\mu (t)= \delta q  ^ \mu+
\deltabar b ^ \mu  $, the action
changes by
\begin{eqnarray}
  \deltabar {\cal A}
= M\int^{t_b}_{t_a} dt \left( g_{\mu \nu }
               \dot{q}^\nu \deltabar \dot{q}^\mu + \frac{1}{2}
              \partial _\mu g_{\lambda \kappa }
             \deltabar q^\mu     \dot{q}^\lambda \dot{q}^\kappa \right).
\label{10.pvar0}\end{eqnarray}
After a partial integration of the $ \delta \dot q$-term
we use (\ref{10.versch}),
(\ref{10.pdelta}), and the identity
$
 \partial _\mu g_{ \nu  \lambda  }
\equiv \Gamma _{\mu \nu  \lambda  }+
\Gamma _{\mu  \lambda  \nu  }$,
which follows directly form the definitions
$g_{ \mu  \nu  }\equiv e^i{}_ \mu  e^i{}_ \nu  $ und
 $\Gamma _{\mu \nu }{}^\lambda  \equiv e_i{}^\lambda
 \partial _\mu e^i{}_\nu $, and obtain
\begin{equation}
 \deltabar {\cal A}
= M\!\!\int^{t_b}_{t_a} dt \bigg[ \!-g_{\mu \nu } \left(
               \ddot{q}^\nu  + \bar \Gamma _{ \lambda  \kappa }{} ^\nu
                \dot{q}^\lambda \dot{q}^\kappa \right)\delta q^ \mu
+\left(g_{ \mu  \nu }
\dot q^ \nu \frac{d}{dt} \deltabar b ^ \mu +
\Gamma _{ \mu  \lambda   \kappa  }
\deltabar b^ \mu \dot{q}^ \lambda \dot{q} ^ \kappa
\right)\bigg]\!.
\label{10.pvar}\end{equation}

To derive the equation of motion we first
vary the action
in a space without torsion.
Then
$\deltabar b^\mu (t) \equiv 0$, and we obtain
\begin{equation}
   \deltabar{\cal A}= \delta {\cal A} = -
M\int _{t_a}^{t_b}dt
g_{\mu \nu }(\ddot{q}^\nu + \bar{\Gamma }_{\lambda \kappa }{}^\nu
        \dot{q}^\lambda \dot{q}^\kappa ) q^ \nu .
\label{10.delact}\end{equation}
Thus, the action principle $\deltabar{\cal A}=0$ produces
the equation for the
 geodesics  (\ref{10.9}), which are the correct particle
 trajectories in the absence of torsion.

In the presence of torsion,
$\deltabar b^ \mu \neq 0$, and the equation of motion
receives a contribution
from the second parentheses in
(\ref{10.pvar}).
After inserting
(\ref{10.p83}), the nonlocal
terms proportional to
$\deltabar b^ \mu $ cancel
and
the total nonholonomic variation of the action becomes
\begin{eqnarray}
  \deltabar{\cal A} & = & -M \int _{t_a}^{t_b}dt g_{\mu \nu }\left[
\ddot{q}^\nu +
     \left( \bar{\Gamma }_{\lambda \kappa }{}^\nu +2S^ \nu {} _{\lambda \kappa
}
 \right) \dot{q}^\lambda \dot{q}^\kappa \right] \delta q^ \mu  \nonumber \\
   & = & -M \int _{t_a}^{t_b}dt g_{\mu  \nu }\left( \ddot{q}^\nu  +
          \Gamma _{\lambda \kappa }{}^\nu \dot{q}^\lambda
           \dot{q}^\kappa \right) \delta q^ \mu  .
\label{10.delat}\end{eqnarray}
The second line follows from the first after using the identity
$\bar{\Gamma }_{\lambda \kappa }{}^\nu  = \Gamma _{\{\lambda \kappa \}}{}^\nu
 + 2 S ^\nu {}_{\{\lambda \kappa \}}$.
The curly brackets indicate the symmetrization
of the enclosed indices.
Setting $\deltabar{\cal A}=0$ gives the autoparallels
(\ref{10.21}) as
the equations of motions,
which is what we wanted to show.

\section[Alternative Formulation of Action Principle with Torsion]
{Alternative Formulation of Action Principle \\with Torsion}

The above variational treatment of the action
is still somewhat complicated and
calls for
an simpler
procedure which we are now going to present.\footnote{See
\aut{H. Kleinert} und \aut{A. Pelster},
FU-Berlin preprint, May 1996.}

Let us vary the paths $q^\mu(t)$ in the usual holonomic
way,
i.e.,
with fixed endpoints,
and consider the associated variations
 $ \delta x ^i=e^i{}_\mu (q) \delta q^\mu$
of the cartesian coordinates.
Taking their time derivative $d_t\equiv d/dt$
we find
\begin{equation}
{d _t} \, \delta x^{\,i} = e^{\,i}_{\,\,\,\lambda} ( q )
d_t \delta q^{\,\lambda}
\label{COM1}
+ \partial_{\mu} e^{\,i}_{\,\,\,\lambda} ( q ) \dot{q}^{\,\mu}
 \delta q^{\,\lambda} .
\end{equation}
On the other hand,
we may write the relation
(\ref{10.diffi}) in the form
$d_tx^i=e^i{}_ \mu (q)d_tq^ \mu$
 and vary  this to yield
\begin{equation}
\delta d_t x^{\,i} \, = \, e^{\,i}_{\,\,\,\lambda}
( q )
 \delta\dot q^{\,\lambda}
\label{COM2}
+  \partial_{\mu}
 e^{\,i}_{\,\,\,\lambda} ( q ) \, \dot{q}^{\,\lambda} \, \delta
q^{\,\mu} \, .
\end{equation}
Using now the fact that
time derivatives $ \delta$ and variations $d_t$
commute
for cartesian paths,
\begin{equation}
\delta  {d_t} x^{\,i}
-  {d_t} \delta x^{\,i}
 =  0  ,
\end{equation}
we deduce from (\ref{COM1}) and (\ref{COM2}) that
this is no longer true
in the presence of torsion, where
\begin{equation}
\label{COMMUTE}
\delta d_t q^{\lambda}
- d_t \delta q^{\lambda}
  =
 2 \, S_{\mu\nu}^{\,\,\,\,\,\lambda} ( q ) \,
\dot{q}^{\mu} \, \delta q^{\,\nu} \, .
\end{equation}
In other words, the variations of the
velocities $\dot q^\mu(t)$
no longer coincide with
the time derivatives of the variations
of $  q^\mu(t)$.

This failure to
to commute is responsible for
shifting the trajectory from
geodesics to
autoparallels.
Indeed, let us vary an  action
\begin{eqnarray}
\label{ACTION}
{\cal A}
= \int\limits_{t_a}^{t_b} dt
L
\left( q^{\,\lambda} ( t ), \dot{q}^{\,\lambda} ( t ) \right)
\end{eqnarray}
by $ \delta q^ \lambda(t)$ and impose
(\ref{COMMUTE}), we find
\begin{eqnarray}
\delta {\cal A} =  \int\limits_{t_a}^{t_b}dt
 \left\{ \frac{\partial L}{\partial q^{\lambda}}
 \delta  q^{\lambda}
+ \frac{\partial L}{\partial \dot{q}^{\lambda}}  \frac{d}{d t}  \delta
q^{\lambda} \right.
 \left. +  2 \,S_{\mu\nu}^{\,\,\,\,\,\lambda}
 \frac{\partial L}{\partial \dot{q}^{\lambda} } \,
\dot{q}^{\mu}  \delta q^{\nu} \right\}  .
\label{VARIATION}
\end{eqnarray}
After a partial integration of the second term
using the vanishing $ \delta q^ \lambda(t)$
at the endpoints,
we obtain
the
Euler-Lagrange equation
\begin{eqnarray}
&&\frac{\partial L}{\partial q^{\,\lambda} } -
\frac{d}{d t} \frac{\partial L}{\partial
\dot{q}^{\lambda} }
= 2  S_{\lambda\mu}^{\,\,\,\,\,\nu}
\dot{q}^{\mu} \frac{\partial
L}{\partial \dot{q}^{\nu} }       .
\label{EL}
\end{eqnarray}
This differs from the standard Euler-Lagrange equation
by
an additional contribution due to the torsion tensor.
For the action (\ref{10.3})
we thus obtain the equation of motion
\begin{equation}
M \, \Big[ g^{\lambda \kappa}  \Big( \partial_{\mu}
g_{\nu\kappa}  - \frac{1}{2} \, \partial_{\kappa}
g_{\mu\nu}  \Big)
+  2 S^{\,\lambda}_{\,\,\,\mu\nu}
\Big]\dot{q}^{\,\mu}  \dot{q}^{\nu} =
0 ,
\end{equation}
which is once more Eq.~(\ref{10.21}) for autoparallels.

\section
{Path Integral in Spaces with Curvature and Torsion}
\index{space with curvature and torsion}
We now turn
to the quantum mechanics of a point particle in a general
\ind{metric-affine space}.
Proceeding in analogy with
the earlier treatment in spherical
coordinates, we first consider the path integral in a flat space with
Cartesian coordinates
\begin{equation} \label{10.49}
({\bf x}\,t \vert {\bf x}'t') =
\frac{1}{\sqrt{2\pi i \epsilon\hbar/M}^D}\prod_{n=1}^{N}
\left[
\int_{-\infty}^{\infty} dx_n \right] \prod_{n=1}^{N+1} K_0^\epsilon (\Delta
{\bf x}_n),
\end{equation}
where $K_0^\epsilon (\Delta {\bf x}_n)$ is an
abbreviation for the short-time amplitude
\begin{equation} \label{10.50}
K_0^\epsilon (\Delta {\bf x}_n) \equiv
\langle {\bf x}_n \vert\exp
\left( -\frac{i}{\hbar} \epsilon \hat{H}\right)
\vert {\bf x}_{n-1}\rangle
= \frac{1}{\sqrt{2\pi i \epsilon \hbar/M}^D}
\exp\left[{\frac{i}{\hbar}\frac{M}{2}\frac{(\Delta{\bf x}_n)^2}{\epsilon}}
\right]
\end{equation}
with $\Delta {\bf x}_n \equiv {\bf x}_n -{\bf x}_{n-1},\, {\bf x}
\equiv {\bf x}_{N+1},\,
{\bf x}' \equiv {\bf x}_0$.
A possible external
potential has been
 omitted since this would
contribute  in
an additive way,
uninfluenced by the space geometry.

Our basic postulate is that the path integral
in a general \ind{metric-affine space} should be obtained
by an appropriate
nonholonomic\ins{nonintegrable mapping}\ins{nonholonomic mapping}
transformation of the amplitude (\ref{10.49})
to a \ind{space with curvature and torsion}.

\subsection{Nonholonomic Transformation of the Action}
The short-time action contains the
square distance $(\Delta {\bf x}_n)^2$
which we have to transform to $q$-space.
For an infinitesimal coordinate difference $\Delta {\bf x}_n\approx d{\bf
x}_n$,
the square distance is obviously given by
$(d{\bf x})^2 = g_{\mu\nu} dq^\mu dq^\nu$. For a
finite $\Delta {\bf x}_n$, however,
it is well known that we must expand $(\Delta {\bf x}_n)^2$ up to
the fourth order
in $\Delta {q_n}^\mu =  {q_n}^\mu - {q_{n-1}}^\mu$
to find all terms contributing to
the  relevant order $\epsilon$.

It is important to realize that with the mapping from
$  d x^i$ to $ d q^ \mu $ not being holonomic,
the finite quantity $ \Delta q^ \mu $
is not uniquely determined by $  \Delta x^i$.
A unique relation can only be obtained by
integrating
the functional
relation
(\ref{10.pr2}) along a specific path.
The preferred path is the classical orbit,
i.e., the autoparallel in the $q$-space.
It is characterized by being the image of a straight line in
the $x$-space. There
$\dot x^i(t)=$const and
the orbit has the linear time dependence
\begin{equation}
\Delta x^i(t)=
\dot x^i(t_0)\Delta t,
 \label{10.linear}\end{equation}
where the time $t_0$ can lie anywhere
on the $t$-axis.
Let us choose for $t_0$ the final time in each interval $(t_n,t_{n-1}).$
At that time,
$\dot x^i_n\equiv
\dot x^i(t_n)$
is related to
$\dot q^ \mu _n\equiv
\dot q^ \mu (t_n)$
by
\begin{equation}
\dot x^i_n=e^i{}_ \mu( q _n) \dot q^ \mu _n.
\label{10.arbit}\end{equation}
It is easy to express
$\dot q^ \mu _n$ in terms
of
$
\Delta q^ \mu _n=
 q^ \mu _n- q^ \mu _{n-1}
$
along the classical orbit.
First we expand
$q^ \mu(t _{n-1})$
into a Taylor series
around
$t_n$. Dropping the time arguments, for brevity, we have
\begin{equation} \label{10.66}
\Delta q\equiv q^\lambda - q'^\lambda =
\epsilon\dot{q}^\lambda
-\frac{\epsilon^2}{2!}\ddot{q}^\lambda
+ \frac{\epsilon^3}{3!}
{{\dot{\ddot{q}}}^\lambda}+\dots~,
\end{equation}
where $ \epsilon =t_n-t_{n-1}$ and $\dot q^\lambda ,\ddot q^\lambda ,
\dots ~$ are the time derivatives at the
final time $t_n$. An expansion of this type is referred to as a
 \iem{postpoint expansion}.
Due to the arbitrariness of the choice of the time $t_0$ in
Eq.~(\ref{10.arbit}),
 the expansion can be performed around any other point just as well,
such as $t_{n-1}$ and $\bar t_n =(t_n+t_{n-1})/2$,
giving rise to the so-called \iem{prepoint} or \iem{midpoint}
expansions of $\Delta q$.

Now, the term $\ddot{q}^\lambda$ in (\ref{10.66}) is given by the equation of
motion
(\ref{10.21}) for the autoparallel
\begin{equation} \label{10.67b}
\ddot{q}^\lambda = -
{\Gamma_{\mu\nu}}{}^\lambda \dot{q}^\mu \dot{q}^\nu.
\end{equation}
 A further time derivative determines
\begin{equation} \label{10.68}
 \dot{\ddot{q}}^\lambda = - (\partial_\sigma {{{\Gamma}}_{\mu\nu}}{}^\lambda
- 2 {{{\Gamma}}_{\mu \nu}}{}^\tau{{{\Gamma}}_{\{\sigma\tau\}}}{}^\lambda  )
\dot{q}^\mu  \dot{q}^\nu \dot{q}^\sigma.
\end{equation}
Inserting  these expressions
into (\ref{10.66}) and inverting the expansion,
 we  obtain
$\dot{q}^\lambda$ at the final time $t_n$ expanded in powers of $\Delta q$.
Using (\ref{10.linear}) and (\ref{10.arbit}) we arrive at the
mapping of the finite coordinate differences:
\begin{eqnarray}
&&
\!\!\!\!\!
\!\!\!\!\!
\Delta x^i=
e^i{}_\lambda \dot q^ \lambda  \Delta t \label{10.69}
 \\
&&\!\!\!\!=
e^i{}_\lambda
\left[ \Delta q^\lambda \!-\!
\frac{1}{2!} {{{\Gamma}}_{\mu\nu}}{}^\lambda
\Delta q^\mu \Delta q^\nu
 \!+\! \frac{1}{3!}
(\partial_\sigma {{{\Gamma}}_{\mu\nu}}{}^\lambda
\!+\! {{{\Gamma}}_{\mu\nu}}{}^\tau{{{\Gamma}}_{\{\sigma\tau\}}}{}^\lambda )
\Delta q^\mu \Delta q^\nu \Delta q^\sigma
\!+\! \dots\right]\!,\nonumber
\end{eqnarray}
where $e^i{}_ \lambda$  and $\Gamma _{\mu \nu }{}^\lambda $
are evaluated at the postpoint.
Inserting this into the short-time amplitude
(\ref{10.50}),
 we obtain
\begin{equation} \label{10.55}
K_0^\epsilon(\Delta {\bf x})\! =\! \langle {\bf x}\vert \exp\left(
-\frac{i}{\hbar}
\epsilon \hat{H}\right) \!
\vert {\bf x}-\Delta {\bf x}\rangle\! =\!
\frac{1}{\sqrt{2\pi i\epsilon\hbar/M}^D} \exp \left[\frac{i}{\hbar}
{\cal A}_>^\epsilon (q,q-\Delta q)\right]
\end{equation}
with the short-time \ind{postpoint action}
\begin{eqnarray}
&&
\!\!\!\!
\!\!\!
\!\!\!
{\cal A}_>^\epsilon(q, q-\Delta q) =
\frac{M}{2\epsilon}( \Delta x^i)^2=
\epsilon\frac{M}{2} g_{\mu\nu} \dot q^\mu \dot q^\nu
\nonumber \\
&&\!\!\!\!=
\bigg\{ g_{\mu\nu} \Delta q^\mu \Delta q^\nu -
 {\Gamma}_{\mu\nu\lambda}
\Delta q^\mu \Delta q^\nu \Delta q^\lambda  \label{10.56}
  \\
& &~ +\left[ \frac{1}{3} g_{\mu\tau} (\partial_\kappa
    { \Gamma_{\lambda\nu}}{}^\tau+
    { \Gamma_{\lambda\nu}}{}^\delta  { \Gamma_{\{ \kappa \delta\}}}{}^\tau)
+ \frac{1}{4} { \Gamma_{\lambda \kappa }}{}^\sigma
 \Gamma_{\mu\nu\sigma}\right]
\Delta q^\mu \Delta q^\nu \Delta q^\lambda \Delta q^\kappa+\dots~\bigg\}.
\nonumber
\end{eqnarray}
Separating the affine connection into Christoffel symbol and
torsion, this can also be written as
\begin{eqnarray} \label{10.56b}
&&
\!\!\!
\!\!\!\!\!\!\!
\!\!\!\!\!\!\!\!\!\!\!\!\!\!\!{\cal A}_>^\epsilon(q, q-\Delta q)  =
\frac{M}{2\epsilon}
\bigg\{ g_{\mu\nu} \Delta q^\mu \Delta q^\nu - \bar{\Gamma}_{\mu\nu\lambda}
\Delta q^\mu \Delta q^\nu \Delta q^\lambda \nonumber  \\
& &+\left[ \frac{1}{3} g_{\mu\tau} (\partial_\kappa
    {\bar\Gamma_{\lambda\nu}}{}^\tau+
    {\bar\Gamma_{\lambda\nu}}{}^\delta  {\bar\Gamma_{\delta \kappa }}{}^\tau)
+ \frac{1}{4} {\bar\Gamma_{\lambda \kappa }}{}^\sigma
\bar\Gamma_{\mu\nu\sigma}\right]
\Delta q^\mu \Delta q^\nu \Delta q^\lambda \Delta q^\kappa
\nonumber \\&&~~~~~~~~~~~~~~~+\frac{1}{3}S^ \sigma {}_{    \lambda \kappa
}S_{ \sigma  \mu  \nu }
+\dots \bigg\}.
\end{eqnarray}

Note that
the right-hand side contains
 only quantities \iem{intrinsic} to the $q$-space.
For the systems treated there (which
all lived in a euclidean space
parametrized with curvilinear coordinates),
the present intrinsic result reduces to the previous one.

At this point we observe that
the final
short-time action
(\ref{10.56})
could also have been
introduced
 without any reference to the
flat reference coordinates $x^i$.
Indeed, the same action is obtained by evaluating
the continuous action (\ref{10.3})
for the small time interval $ \Delta t= \epsilon $ along the
classical orbit between the points $q_{n-1}$
and
 $q_{n}$.
Due to the equations of motion
(\ref{10.21}), the Lagrangian
\begin{equation} \label{10.63}
L(q,\dot{q}) = \frac{M}{2}g_{\mu\nu} (q(t)) \,\dot{q}^\mu(t) \dot{q}^\nu(t)
\end{equation}
is independent of time (this is true for
 autoparallels as well as geodesics). The short-time action
\begin{equation} \label{10.64}
{\cal A}^\epsilon (q,q') = \frac{M}{2} \int_{t-\epsilon }^{t} dt
\,g_{\mu\nu} (q(t)) \dot{q}^\mu(t) \dot{q}^\nu(t)
\end{equation}
can therefore be written in either of the three forms
\begin{equation} \label{10.65}
{\cal A}^\epsilon = \frac{M}{2}\epsilon  g_{\mu\nu} (q)\dot{q}^\mu  \dot{q}^\nu
= \frac{M}{2}\epsilon  g_{\mu\nu} (q')\dot{q}'{}^\mu  \dot{q}'{}^\nu
= \frac{M}{2}\epsilon  g_{\mu\nu} (\bar{q}) {\dot{\bar{q}}}^\mu
{\dot{\bar{q}}}^\nu ,
\end{equation}
where ${q}{}^\mu,{q}'{}^\mu, {{\bar{q}}}^\mu$
are the coordinates at the final time $t_{n}$,
the initial time $t_{n-1}$,
and the average time
$(t_n+t_{n-1})/2$, respectively.
The first expression obviously coincides with
(\ref{10.65}).  The others can be used
as a starting point for deriving
equivalent
prepoint or midpoint actions.\ins{midpoint}\ins{prepoint action}
The prepoint action ${\cal A}_<^\epsilon$
arises from
the postpoint one ${\cal A}_>^\epsilon$ by exchanging
$ \Delta q$ by $- \Delta q$ and the postpoint coefficients by the prepoint
ones.
The midpoint action
has the most simple-looking appearance:
\begin{eqnarray} \label{10.59}
\lefteqn{\!\!\!\!\!\!\!\!\!\!~
\bar{\cal A}^\epsilon (\bar{q} + \frac{\Delta q}{2},
\bar{q} - \frac{\Delta q}{2}) =}\\
\!\!\!\!
\!\!\!\!
&&\!\! \frac{M}{2 \epsilon }
\left[ {g}_{\mu\nu}(\bar q) \Delta q^\mu \Delta q^\nu
\!+\!\frac{1}{12} g_{\kappa\tau} (\partial_\lambda {\Gamma_{\mu\nu}}^\tau
\!+ \!{\Gamma_{\mu\nu}}^\delta {\Gamma_{\{\lambda\delta\}}}^\tau )
 \Delta q^\mu \Delta q^\nu\Delta q^\lambda
\Delta q^\kappa+\dots \right],\nonumber
\end{eqnarray}
where
the affine connection can be evaluated at any point
in the interval $(t_{n-1},t_{n})$. The precise
position is irrelevant to the
amplitude producing only changes beyond the relevant order epsilon.

We have found the postpoint action most
useful since it gives ready access to the time evolution of amplitudes,
as will be seen below. The prepoint action is completely equivalent to it
and useful if one wants to describe the time evolution
backwards. Some authors favor the midpoint action
 because of its
symmetry and intimate relation to
an ordering prescription in operator quantum mechanics which was
advocated by
\aut{H.~Weyl}.
This prescription is, however,
 only of historic interest since it
does not lead to the correct physics.
In the following, the action ${\cal A}^\epsilon $ without
subscript will  always denote the preferred postpoint
expression (\ref{10.56}):
\begin{equation} \label{10.62}
{\cal A}^\epsilon \equiv {\cal A}_>^\epsilon(q, q-\Delta q) .
\end{equation}
%

\subsection{The Measure of Path Integration}

\index{measure, path}\index{path measure}
We now turn to the integration
measure in the Cartesian path integral
(\ref{10.49})
$$
\frac{1}{\sqrt{2\pi i\epsilon \hbar/M}^D}
\prod_{n=1}^{N} d^Dx_n .
$$
This has to be transformed
to the general metric-affine space.\index{metric-affine space, path measure}
We imagine evaluating the path integral
starting out from the latest time and performing successively the
integrations over $x_N , x_{N-1},\dots~$,
 i.e., in each short-time amplitude we integrate over the earlier
position coordinate,
 the prepoint
coordinate.  For the purpose of this discussion, we
relabel the product
$\prod_{n=1}^{N} d^Dx_n^i$ by $\prod_{n=2}^{N+1} dx_{n-1}^i$,
so that the integration in
each time slice  $(t_n,t_{n-1})$  with $n=N+1,N,\dots $
runs over $ dx_{n-1}^i$.

In a flat space parametrized with
curvilinear coordinates,
the transformation of the integrals
over $d^Dx_{n-1}^i$ into those over $d^Dq_{n-1}^\mu$ is obvious:
\begin{equation} \label{10.77}
\prod_{n=2}^{N+1} \int d^Dx_{n-1}^i =
\prod_{n=2}^{N+1}\left
\{\int d^Dq_{n-1}^\mu~\det\left[e_\mu^i
(q_{n-1})\right] \right\}.
\end{equation}
The determinant of ${e^i}_\mu$ is the square root of
the determinant of the metric $g_{\mu\nu}$:
\begin{equation} \label{10.78}
\det({e^i}_\mu) = \sqrt{\det g_{\mu\nu}(q)} \equiv \sqrt{g(q)},
\end{equation}
and
the measure\index{measure, path}\index{path measure}
may be rewritten as
\begin{equation} \label{10.79}
\prod_{n=2}^{N+1} \int d^Dx_{n-1}^i = \prod_{n=2}^{N+1}\left
[\int d^Dq_{n-1}^\mu~\sqrt{ g(q_{n-1})}
 \right].
\end{equation}
This expression is not directly applicable.
When trying to do the
$d^Dq^\mu_{n-1}$-integrations
successively, starting from the final integration over $dq_N^\mu $,
the integration variable  $q_{n-1} $
appears for each $n$
in
the
argument  of
$\det\left[e_\mu^i
(q_{n-1})\right]$ or $ g_{ \mu  \nu }(q_{n-1})$.
To make this $q_{n-1}$-dependence explicit,
 we
expand in the measure (\ref{10.77})
 $e_\mu^i(q_{n-1})=e^i{}_\mu (q_n-\Delta q_n)$
around the postpoint $q_n $ into
powers of $\Delta q_n$. This gives
\begin{equation} \label{10.80}
dx^i = e_\mu^i (q-\Delta q) dq^\mu = e^i_\mu dq^\mu -{e^i}_{\mu,\nu}
dq^\mu \Delta q^\nu + \frac{1}{2} {e^i}_{\mu,\nu\lambda} dq^\mu \Delta q^\nu
\Delta q^\lambda + \dots~,
\end{equation}
 omitting, as before, the
subscripts of $q_n$ and $\Delta q_n$. Thus the
Jacobian of the coordinate transformation from $dx^i$ to $dq^\mu$
is
\begin{equation} \label{10.81}
J_0 = \det ({e^i}_ \kappa  )
{}~\det \left[{\delta^\kappa}_\mu -
{e_i}^\kappa {e^i}_{\mu, \nu} \Delta q^\nu + \frac{1}{2}
{e_i}^\kappa {e^i}_{\mu,\nu\lambda} \Delta q^\nu \Delta q^\lambda \right],
\end{equation}
giving the relation between the infinitesimal integration
volumes $d^Dx^i$ and $d^Dq^\mu$:
\begin{equation} \label{10.82}
\prod_{n=2}^{N+1} \int d^Dx_{n-1}^i =
\prod_{n=2}^{N+1}\left
\{\int d^Dq_{n-1}^\mu
\,J_{0n}\right\}.
\end{equation}
The well-known expansion  formula
\begin{equation} \label{10.83}
\det(1+B) =\exp \mbox{tr}\log (1+B) =\exp
\mbox{tr} (B-B^2/2+B^3/3-\dots)
\end{equation}
allows us now to rewrite $J_0$ as
\begin{equation} \label{10.84}
J_0 = \det (e^i{}_ \kappa ) \exp
\left( \frac{i}{\hbar}{\cal A}_{J_0}^\epsilon  \right),
\end{equation}
with the determinant $\det (e^i_\mu)=\sqrt{g(q)}$ evaluated at
the postpoint.
This equation
defines an effective action associated with the
Jacobian,\index{Jacobian action}\index{action, Jacobian}
for which we obtain
the expansion
\begin{equation} \label{10.85}
\!\frac{i}{\hbar} {\cal A}^\epsilon _{J_0} = -{e_i}^\kappa {e^i}_{\kappa,\mu}
\Delta q^\mu\! + \!\frac{1}{2} \left [{e_i}^\mu e^i_{ \mu,\nu\lambda }
\!- {e_i}^\mu e^i{}_{ \kappa,\nu }
{e_j}^\kappa {e^j}_{\mu,\lambda}\right ]
\Delta q^\nu \Delta q^\lambda +\dots~.
\end{equation}
To express this in terms of the affine connection,
we use (\ref{10.19}) and
derive the relations
\begin{eqnarray} \label{10.53}
\frac{1}{4} e_{i\nu ,\mu} {e^i}_{\kappa , \lambda} &=&
\frac{1}{4} {e_i}^\sigma {e^i}_{\nu ,\mu} e_{j\sigma}
{e^j}_{\kappa ,\lambda} = \frac{1}{4} {\Gamma_{\mu\nu}}^\sigma,
\Gamma_{\lambda\kappa\sigma}
\\
 \label{10.54}
\frac{1}{3} e_{i\mu} {e^i}_{\nu ,\lambda\kappa} & = &
     \frac{1}{3} g_{\mu\tau} [\partial_\kappa
     ({e_i}^\tau {e^i}_{\nu ,\lambda}) -
     e^{i\sigma} {e^i}_{\nu ,\lambda} {e^j}_\sigma
     {e^{j\tau}}_{,\kappa}]    \nonumber \\
 & = &  \frac{1}{3} g_{\mu\tau} (\partial_\kappa {\Gamma_{\lambda\nu}}^\tau
   + {\Gamma_{\lambda\nu}}^\sigma  {\Gamma_{\kappa\sigma}}^\tau).
\end{eqnarray}
With these, the Jacobian action
 becomes
\begin{eqnarray} \label{10.86}
\frac{i}{\hbar}{\cal A}^\epsilon _{J_0}
&=&-\Gamma_{\mu\nu}{}^\nu \Delta q^\mu
 +  \frac{1}{2} \partial_{\mu}
\Gamma_{\nu\kappa }{}^\kappa
 \Delta q^\nu \Delta q^\mu + \dots~.
\end{eqnarray}
The same result would, of course, be obtained by writing the Jacobian in
accordance with (\ref{10.79})
as
\begin{equation}
J_0=\sqrt{g(q- \Delta q)} ,
\label{}\end{equation}
which leads to the alternative formula for the Jacobian action
\begin{equation}
\exp\left(\frac{i}{\hbar}{\cal A}^\epsilon _{J_0}\right)= \frac{\sqrt{g(q-
\Delta q)}}{
\sqrt{g(q)}}.
\label{}\end{equation}
An expansion in powers of $ \Delta q$ gives
\begin{eqnarray} \label{11.42cb}
{}~~\!\!\! \exp\left(\frac{i}{\hbar}{\cal{A}}^\epsilon _{\bar J_0}\right)
\! =  \!1 \! -\frac{1}{\sqrt{g(q)}}\sqrt{g(q)}_{,\mu}
\Delta q^{\mu} \! + \frac{1}{2\sqrt{g(q)}} \sqrt{g(q)}_{,\mu\nu}\Delta
q^{\mu} \Delta q^{\nu} \! +\!\dots~.\nonumber \\
\end{eqnarray}
Using the formula
\begin{eqnarray} \label{11.24a}
\frac{1}{\sqrt{g}} \partial_{\mu}\sqrt{g}  = \frac{1}{2}
 g^{\sigma \tau }\partial _\mu
g_{\sigma \tau }= \bar{\Gamma}_{\mu\nu}^{~~\,\nu},
\end{eqnarray}
this becomes
\begin{eqnarray}
&&\!\!\! \exp\left(\frac{i}{\hbar}{\cal{A}}^\epsilon _{\bar J_0}\right)
  =  1 - {\bar{\Gamma}_{\mu\nu}}{}^{\nu} \Delta q^{\mu} +
 \frac{1}{2} (\partial_{\mu} {\bar{\Gamma}_{\nu\lambda}}{}^{\lambda}
{+\bar{\Gamma}_{\mu\sigma}}^{\sigma}
{\bar{\Gamma}_{\nu\lambda}}{}^{\lambda})
\Delta q^{\mu} \Delta q^{\nu}+\dots,\nonumber \\~~   \label{11.42bb}
\end{eqnarray}
so that
\begin{equation} \label{11.42c}
\frac{i}{\hbar}{\cal{A}}^\epsilon _{\bar J_0}
  =  - {\bar{\Gamma}_{\mu\nu}}{}^{\nu} \Delta q^{\mu} +
 \frac{1}{2} \partial_{\mu} {\bar{\Gamma}_{\nu\lambda}}{}^{\lambda}
\Delta q^{\mu} \Delta q^{\nu}+\dots~.
\end{equation}
In a space without torsion where
$\bar {\Gamma}_{\mu\nu}^\lambda \equiv {\Gamma }_{\mu\nu }{}^{\lambda}$,
the Jacobian actions (\ref{10.86})
and (\ref{11.42c})
are trivially equal to each other.
But the equality holds also in the presence of torsion.
Indeed, when
inserting the decomposition (\ref{10.26b}),
${\Gamma_{\mu\nu}}^\lambda ={ \bar\Gamma }_{\mu\nu }^{\;\;\;\;\lambda} +
{K_{\mu\nu}}^\lambda,$ into
 (\ref{10.86}),
the contortion tensor
drops out
since it is antisymmetric in the last two indices
and these are contracted in both expressions.

In terms of ${\cal A}_{J_{0n}}^\epsilon$,
we can rewrite the transformed measure
(\ref{10.77})
in the more useful form
\begin{equation} \label{10.77da}
\prod_{n=2}^{N+1} \int d^Dx_{n-1}^i =
\prod_{n=2}^{N+1}\left\{
\int d^Dq_{n-1}^\mu~\det\left[e_\mu^i(q_{n})\right]
 \exp\left( \frac{i}{\hbar}{\cal A}_{J_{0n}}^\epsilon  \right)
\right\}.
\end{equation}

In a flat space parametrized in terms of curvilinear coordinates,
the two sides of
(\ref{10.77}) and
(\ref{10.77da})
 are related by an ordinary coordinate transformation,
and the right-hand side
gives the correct measure
for a
time-sliced path integral.
In
a general \ind{metric-affine space}, however,
this is no longer true.
Since the mapping
$dx^i\rightarrow dq^\mu
$
is nonholonomic,
there are in principle infinitely many ways of transforming
the path integral measure from Cartesian coordinates
to a noneuclidean space.
Among these, there exists
a preferred mapping which
leads to the correct quantum-mechanical amplitude
in all known physical systems.
It is this mapping which led to the
correct solution of the path integral of the hydrogen atom \cite{dk}.

The clue for finding the correct mapping
is offered by
an
unesthetic
feature of
Eq.~(\ref{10.80}):
The expansion
contains both differentials $dq^\mu$ and differences $\Delta q^\mu$.
This is somehow inconsistent.
When
time-slicing the path integral, the
differentials
$dq^\mu$ in the action are increased to finite differences $\Delta q^\mu$.
Consequently, the
differentials in the
measure should\index{measure, path}\index{path measure}
also become differences. A
relation such as (\ref{10.80})
containing simultaneously
differences and differentials should not occur.

It is easy to achieve this goal
by changing the starting point of the nonholonomic mapping
and rewriting
the initial flat space path integral
(\ref{10.49}) as
\begin{equation} \label{10.49b}
({\bf x}\,t \vert {\bf x}'t') =
\frac{1}{\sqrt{2\pi i \epsilon\hbar/M}^D}\prod_{n=1}^{N}
\left[
\int_{-\infty}^{\infty} d \Delta x_n \right] \prod_{n=1}^{N+1} K_0^\epsilon
(\Delta
{\bf x}_n).
\end{equation}
Since $x_n$ are
Cartesian coordinates, the  measures\index{measure, path}\index{path measure}
of integration in the
time-sliced expressions  (\ref{10.49}) and (\ref{10.49b})
are certainly identical:
\begin{equation} \label{10.87}
\prod_{n=1}^{N} d^D x_n \equiv  \prod_{n=2}^{N+1} d^D \Delta x_n.
\end{equation}
Their images under a nonholonomic mapping, however,
are different so that the initial form of the time-sliced
path integral is a matter of choice.
The initial form (\ref{10.49b})
has the obvious advantage
that the integration variables are
precisely the quantities  $\Delta x^i_n$
which occur in the
short-time amplitude $K_0^ \epsilon ( \Delta x_n)$.

Under a nonholonomic transformation,
the right-hand side of Eq.~(\ref{10.87})
leads to the integral measure
in a general \ind{metric-affine space}
\begin{equation} \label{10.77b}
\prod_{n=2}^{N+1} \int d^D\Delta x_{n} \rightarrow
\prod_{n=2}^{N+1} \left[\int d^D\Delta q_{n}\,J_n\right],
\end{equation}
 with the
Jacobian following from (\ref{10.69}) (omitting $n$)
\begin{eqnarray} \label{10.89ex}
J&\!\! =&\! \!\frac{\partial(\Delta x)}{\partial(\Delta q)} \\
&\!\!=&\!     \!
\det(e^i{}_\kappa)\,\det\!\!\left[
 \delta  _\mu {}^ \lambda \!-\!
 {{{\Gamma}}_{\{\mu\nu\}}}{\!}^\lambda
 \Delta q^\nu
 \!+\! \frac{1}{2}
(\partial_\sigma {{{\Gamma}}_{\mu\nu}}{\!}^\lambda
\!+ \!\Gamma_{\{\mu\nu}{}^\tau
\Gamma_{ \{\tau|\sigma \}\} }{\!}^\lambda)
  \Delta q^\nu \Delta q^\sigma
\!+\! \dots\right]\hspace{-2pt}. \nonumber
\end{eqnarray}
In a space with curvature and torsion, the measure on the
right-hand side of
(\ref{10.77b}) replaces the flat-space measure
on the right-hand side of (\ref{10.79}).
The curly double brackets around the indices $ \nu,  \kappa , \sigma , \mu$
indicate a symmetrization
in $\tau $ and $ \sigma $ followed by a symmetrization
in $ \mu  ,\nu$, and  $\sigma $.
With the help of formula (\ref{10.83})
we now calculate
the  Jacobian action\index{Jacobian action}\index{action, Jacobian}
%
%
%
\begin{eqnarray} \!\!\!\!\frac{i}{\hbar}{\cal A}^\epsilon _{J}
&=&-\Gamma_{\{\mu\nu\}}{}^\mu \Delta q^\nu\label{10.91}
 \\
 &&+  \frac{1}{2} \left[\partial_{\{\mu}
\Gamma_{\nu\kappa\} }{}^\kappa + {\Gamma_{\{ \nu\kappa}}^\sigma
\Gamma_{ \{ \sigma|\mu \}\}}{}^\kappa - {\Gamma_{\{ \nu\kappa\} }}^\sigma
{\Gamma_{\{\sigma \mu\} }}^\kappa \right]
\Delta q^\nu \Delta q^\mu + \dots~.
\nonumber
\end{eqnarray}
This expression differs from the
earlier Jacobian action (\ref{10.86}) by the symmetrization symbols.
Dropping them, the two expressions
coincide.
This is allowed if $q^ \mu $ are
curvilinear coordinates in
a flat space.
Since then the
transformation functions
$x^i(q)$ and their first derivatives $\partial_\mu x^i(q)$
are integrable and possess commuting derivatives, the two
Jacobian actions (\ref{10.86}) and (\ref{10.91})
are identical.

There is a further good reason
for choosing
(\ref{10.87}) as a starting point for the nonholonomic transformation
of the measure.\index{measure, path}\index{path measure}
According to
Huygens' principle of wave optics, each point of a wave front
is a center of a new spherical wave propagating from that
point. Therefore, in a
 \ind{time-sliced path integral},
the differences
$\Delta x_n^i$ play a more fundamental role than the
coordinates themselves.  Intimately related to this is the
observation that in the canonical form, a short-time piece
of the action reads
\begin{equation} \label{10.92}
\int \frac{dp_n}{2\pi\hbar} \exp \left[
\frac{i}{\hbar} p_n (x_n-x_{n-1}) - \frac{ip_n^2}{2M\hbar}t\right].\nonumber
\end{equation}
Each momentum is associated with a coordinate difference
$\Delta x_n \equiv x_n~-~x_{n-1}$. Thus, we should expect
the spatial integrations conjugate to $p_n$ to run over the
coordinate differences $\Delta x_n = x_n -x_{n-1}$~~rather than
the coordinates $x_n$ themselves, which
makes the important difference
in the subsequent nonholonomic coordinate transformation.

We are thus led to postulate the following
\index{time-sliced path integral}
time-sliced path integral
in $q$-space:
\begin{eqnarray} \label{10.93}
\lefteqn{\!\!\!\!\!\!\!\!\!\!\!\!\!\!\!\!\!\!\!\!\!\!\!
\!\!\!\!\!\!\!\!\!\!\!\!\!\!\!\!\!\!\!
\langle q\vert \exp\left[ - \frac{i}{\hbar}(t-t')\hat{H}\right]\vert q'
\rangle =
\frac{1}{\sqrt{2\pi i\hbar\epsilon/M}^D}\prod_{n=2}^{N+1}\left[
\int{d^D \Delta q_n}
 \frac{\sqrt{g(q_n)}}{\sqrt{2\pi i\epsilon\hbar/M}^D}\right]
}\nonumber \\
&&~~~~~~~~~~~~~
\times \exp \left[ \frac{i}{\hbar} \sum_{n=1}^{N+1}({\cal A}^\epsilon
 +{\cal A}^\epsilon _{J})\right],
\end{eqnarray}
where the integrals over $\Delta q_n$ may
be performed successively from $n=N$
down to $n=1$.

Let us emphasize that this expression has not been {\em derived} from
the flat space path integral. It is the result of a
specific new \iem{quantum equivalence principle}
which rules how a flat space path integral behaves
under nonholonomic coordinate transformations.

It is useful to reexpress our result
in a different form which clarifies best the relation with
the
naively expected measure of path integration
\index{measure, path}\index{path measure}
(\ref{10.79}), the product of integrals
\begin{equation} \label{10.94}
\!\!\!\!\!\!\!\!\!
\!\!\!\!\!\!\!\!\prod_{n=1}^{N} \int d^Dx_{n} =
\prod_{n=1}^{N}\left[
\int d^Dq_{n}\,\sqrt{ g(q_{n})}
 \right]                               .
\end{equation}
The measure in (\ref{10.93})
can be expressed in terms of (\ref{10.94}) as
\begin{eqnarray} \label{}
\prod_{n=2}^{N+1}  \left[
\int{d^D \Delta q_n}
\sqrt{  g(q_{n})}\right]   =
  \prod_{n=1}^{N}\left
[\int d^Dq_{n}\,\sqrt{ g(q_{n})}
e^{-i {\cal A}^\epsilon _{J_0}/\hbar }
 \right].\nonumber
\end{eqnarray}
The corresponding expression for the entire
time-sliced path
integral\index{measure, path}\index{path measure}
(\ref{10.93})
in the \ind{metric-affine space} reads
\begin{eqnarray} \label{10.93b}
\lefteqn{\!\!\!\!\!\!\!\!\!\!\!\!\!\!\!\!\!\!\!\!\!\!\!\!\!\!\!
\!\!\!\!\!\!\!\!\!\!\!\!\!\!\!\!\!\!\!
\langle q\vert \exp\left[ - \frac{i}{\hbar}(t-t')\hat{H}\right]\vert q'
\rangle =
\frac{1}{\sqrt{2\pi i\hbar\epsilon/M}^D}
 \prod_{n=1}^{N}\left
[\int d^Dq_{n}\frac{\sqrt{ g(q_{n})}}
{\sqrt{2\pi i\hbar\epsilon/M}^D}
  \right]
   }\nonumber \\
&&~~~~~~~
\times \exp \left[ \frac{i}{\hbar} \sum_{n=1}^{N+1}({\cal A}^\epsilon
+\Delta  {\cal A}^\epsilon _{J}) \right],
\end{eqnarray}
where
$ \Delta {\cal A}^\epsilon _{J}$ is the difference between the correct
and
the wrong Jacobian actions in Eqs.~(\ref{10.86}) and (\ref{10.91}):
\begin{eqnarray} \label{10.96}
\Delta    {\cal A}^\epsilon _{J}\equiv  {\cal A}^\epsilon _{J} - {\cal
A}^\epsilon _{J_0}.
\end{eqnarray}

In the absence of torsion where $
\Gamma _{\{\mu \nu \}}{}^\lambda ={\bar\Gamma} _{\mu \nu }{}^\lambda $,
this simplifies to
\begin{eqnarray} \label{10.97}
\frac{i}{\hbar}\Delta {\cal A}^\epsilon _{J}
=\frac{1 }{6}
\bar R_{\mu \nu }\Delta q^\mu\Delta q^\nu ,
\end{eqnarray}
where $ \bar R_{\mu \nu }$ is the Ricci tensor associated with
the Riemann curvature
tensor, i.e., the contraction (\ref{10.36}) of the Riemann
curvature tensor associated with
the Christoffel symbol $\bar \Gamma _{\mu \nu }{}^\lambda $.

Being quadratic in $ \Delta q$, the effect of the additional
action can easily be evaluated perturbatively using the methods
explained in Chapter~8, according to which
$ \Delta q^\mu \Delta q^\nu $ may be replaced
by its lowest order expectation
\begin{eqnarray}
 \langle \Delta q^\mu \Delta q^\nu \rangle _0
=i\epsilon \hbar g^{\mu \nu }(q)/M.
\nonumber
\end{eqnarray}
Then $ \Delta {\cal A}^\epsilon _{J}$ yields
the additional \ind{effective potential}
\begin{equation} \label{10.98}
V_{\rm eff}
=-\frac{\hbar ^2}{6M}\bar R,
\end{equation}
where $ \bar R$ is the Riemann curvature scalar.
By including this potential in the action, the
path integral
\index{measure, path}
\index{path measure}
in a curved space can be
written down in the naive form (\ref{10.94})
 as follows:
\begin{eqnarray} \label{10.99}
\lefteqn{~~\!\!\!\!\!\!\!\!\!\!\!\!\!\!\!\!\!\!\!\!\!\!
\!\!\!\!\!\!\!\!\!\!\!\!\!\!\!\!\!\!\!\!\!\!\!
\langle q\vert \exp\left[ - \frac{i}{\hbar}(t-t')\hat{H}\right]\vert q'
\rangle =
\frac{1}{\sqrt{2\pi i\hbar\epsilon/M}^D}
\prod_{n=1}^{N}\left[
\int{d^D q_{n}}
 \frac{\sqrt{g(q_{n})}}{\sqrt{2\pi i\epsilon\hbar/M}^D} \right]
}\nonumber \\
&&~~~~~~ ~~~~~~
\times \exp \left[ \frac{i}{\hbar} \sum_{n=1}^{N+1}
({\cal A}^\epsilon +\epsilon V_{\eff})\right].
\end{eqnarray}
The integrals over $ q_{n}$ are conveniently
performed successively downwards
over $\Delta q_{n+1}=q_{n+1}-q_n$ at fixed $ q_{n+1}$.
 The weights $ \sqrt{ g(q_{n})}=\sqrt{ g(q_{n+1}-\Delta q_{n+1})}$
require a postpoint expansion  leading to the naive
Jacobian $ J_0$ of (\ref{10.81})
and the \ind{Jacobian action} $
{\cal A}^\epsilon _{J_0}$ of Eq.~(\ref{10.86}).

It goes without saying that the path integral (\ref{10.99})
also has a phase space version.
It is obtained by
omitting all $(M/2\epsilon )(\Delta q_n)^2$
terms in the short-time actions
${\cal A}^\epsilon$ and extending the multiple integral
by the product of momentum integrals
\begin{eqnarray}
\prod _{n=1}^{N+1}\left[ \frac{dp_n}{2\pi \hbar \sqrt{g(q_n)} } \right]
e^{(i/\hbar )\sum _{n=1}^{N+1}\left[
p_{n\mu }\Delta q^\mu -\epsilon\frac{1}{2M}g^{\mu \nu }(q_n)
p_{n\mu }p_{n\nu }
\right]}
{}.
\label{}\end{eqnarray}
When using this expression, all problems which were
encountered  in the literature
with canonical transformations of path integrals
disappear.

\section{Conclusion}
It appears as though the new variational
and quantum equivalence principles
constitute the proper
basis for
a correct extension of our physical laws into
geometries with torsion.
The evidences at present are:

\begin{itemize}

\item
The solution of the path integral
of the Coulomb problem can be found without
undesirable time-slicing corrections \cite{PI}.
\item
The Euler-Lagrange equations of a spinning top
can be derived in the rotating
body-fixed frame of references by extremizing the kinetic
action \cite{fk2}.
\item
The bosonization
\cite{cqf,gor,gl,legg,tom,pair,hs,witten,cm,kr} of Fermi theories
is possible \cite{boz} without errors in the energy spectrum.
\end{itemize}

Fig. 1: {Crystal with dislocation and disclination generated by nonholonomic
coordinate transformations from an ideal crystal.
Geometrically, the former transformation introduces torsion, the latter
curvature.~\\~\\
Fig. 2: Images
under a holonomic and a nonholonomic mapping
of a fundamental path variation.
In the holonomic case,
the paths $x(t)$ and $x(t)+\delta x(t)$  in (a)
turn into the
paths $q(t)$ and $q(t) + \delta q(t)$
in (b). In the
nonholonomic case with $S_ \mu {}^{\nu\lambda  } \neq 0$,
they go over into
$q(t)$ and $q(t)+\deltabar q(t)$
shown in (c) with a \ind{closure failure} $b^ \mu $ at $t_b$  analogous
 to the Burgers vector $b^ \mu$ in a solid with dislocations.}

\begin{figure}[tbhp]
~\\~\\
$\!\!\!\!\!\!\!\!\!\!$\input disloccl.te
\caption{}
\label{}\end{figure}
%
\begin{figure}[tbh]
{}\input nonholm3.te
 \label{10.pnonh}
\caption{}
\end{figure}
%
%

\end{document}